# *Ab initio* modeling of the energy landscape for screw dislocations in body-centered cubic high-entropy alloys

**Sheng Yin**[1,2], **Jun Ding**[2], **Mark Asta**[1,2*] **and Robert O. Ritchie**[1,2*]

[1]Department of Materials Science & Engineering, University of California, Berkeley, CA 94720, USA

[2]Materials Sciences Division, Lawrence Berkeley National Laboratory, Berkeley, CA 94720, USA

*To whom correspondence may be addressed; email: mdasta@berkeley.edu or roritchie@lbl.gov

**ABSTRACT AND KEYWORKDS**

In traditional body-centered cubic (*bcc*) metals, the core properties of screw dislocations play a critical role in plastic deformation at low temperatures. Recently, much attention has been focused on refractory high-entropy alloys (RHEAs), which also possess *bcc* crystal structures. However, unlike face-centered cubic high-entropy alloys (HEAs), there have been far fewer investigations on *bcc* HEAs, specifically on the possible effects of chemical short-range order (SRO) in these multiple principal element alloys on dislocation mobility. Here, using density functional theory, we investigate the distribution of dislocation core properties in MoNbTaW RHEAs alloys, and how they are influenced by SRO. The average values of the core energies in the RHEA are found to be larger than those in the corresponding pure constituent *bcc* metals, and are relatively insensitive to the degree of SRO. However, the presence of SRO is shown to have a large effect on narrowing the distribution of dislocation core energies and decreasing the spatial heterogeneity of dislocation core energies in the RHEA. It is argued that the consequences for the mechanical behavior of HEAs is a change in the energy landscape of the dislocations which would likely heterogeneously inhibit their motion.

**Keywords:** Refractory high-entropy alloys; screw dislocations; Peierls potential; local chemical ordering

**INTRODUCTION**

Previous investigation of the fundamentals of deformation in body-centered cubic (*bcc*) transition metals have revealed that the core properties of the ½<111> screw dislocations play an essential role in their plasticity,[1] especially at low temperatures where the deformation is thermally activated through the kink-pair nucleation mechanism,[2] and expected to be strongly temperature-dependent. The high lattice friction associated with such screw dislocation motion is a result of nonplanar core structure[1,3] and related to the height of the Peierls potential.[4]

Due to the importance for plastic deformation, extensive atomistic simulation studies have been devoted to computing core structures and corresponding mobilities of screw dislocations in *bcc* transition metals.[3,5,6,7,8] In these studies one of the significant challenges has been the variation in properties derived from different models for the interatomic potentials. For example, early studies based on classical potential models often predicted a metastable split core structure,[9,10,11] which leads to a camel-hump shape in the Peierls potential. Later density functional theory (DFT) calculations produced symmetric and compact dislocation cores in Mo, Ta and Fe;[12,13,14,15,16] similar compact cores have been found in other *bcc* transition metals, such as W, Nb and V.[17,18] In DFT studies of the energy landscape of screw dislocations in *bcc* transition metals,[18,19,20] it was found that non-degenerate cores lead to a single humped curve in the Peierls potential, implying that the split core structure might not be metastable. Alloying effects on the Peierls potential of W have also been explored[21]. Recently developed machine learning based potentials[22,23,24] and new embedded atom method (EAM) potentials that consider quantum effects on lattice vibrations[25] and extra constraints[26] all lead to predictions of a single humped curve in the Peierls potential. Due to the dependence of the results for screw dislocations in *bcc* transition metals on the model for interatomic bonding, DFT-based approaches are of interest to provide benchmarks for subsequent modeling at higher scales.

During the past fifteen years, a new class of alloys known as high-entropy alloys (HEAs)[27,28] has drawn extensive research interest. These alloys involve multiple principal elements (typically five) in nominally equimolar ratios, and were originally presumed to crystallize as a single-phase solid solution. As a new class of structural materials, some types of HEAs, in particular the CrCoNi-based alloys, have been shown to possess exceptional damage tolerance and improved strength at cryogenic temperatures.[29,30] Theoretically, mechanistic, first-principles-based predictive theories for the temperature-, composition-, and strain-rate-dependence of the plastic yield strength have been developed and applied to such face-centered cubic (*fcc*) alloys.[31,32,33] Indeed, most HEA research to date has been focused on these *fcc* "Cantor-type" alloys,[34,35] whereas a second distinct family of HEAs, comprising mostly refractory elements, has been far less studied. Such refractory high-entropy alloys (RHEAs), which are sometimes termed Senkov alloys,[36,37] invariably crystallize in *bcc* solid-solution phases that have been designed for elevated temperature applications.[38]

For example, RHEAs such as MoNbTaW with single-phase *bcc* crystal structures have been produced by vacuum arc melting[37] or direct metal deposition[39] with exceptional microhardness[36] as well as excellent compression yield strength and good ductility at high temperatures.[37] Transmission electron microscopy (TEM) studies on RHEAs have shown a dominant role of screw dislocations with increasing plastic strain,[40,41] similar to traditional *bcc* metals. Additionally, strong intrinsic lattice resistance has been found in certain RHEAs.[41,42] To model such behavior, molecular dynamics (MD) simulations have been used to study dislocation behavior in *bcc* RHEAs.[43] For example, screw dislocation core structures in NbTiZr, $Nb_{1.5}TiZr_{0.5}$ and $Nb_{0.5}TiZr_{1.5}$ alloys were recently explored using MD simulations, and significant core structure variation was found along the dislocation line.[44] Recent theory has revealed the potential importance of edge dislocations in controlling the strength of *bcc* HEAs at high temperatures[45] and the correlation between atomic distortions and the yield strengths of HEAs.[46] However, there are still only very limited studies on the deformation behavior of this new class of *bcc* alloys, as compared to single-phase *bcc* transition metals.

Another important aspect of HEAs is the presence of local chemical short-range order (SRO). Although these alloys can be described as "topologically ordered yet chemically disordered", the local chemical environments are unlikely to be characterized by a perfectly random distribution for every atomic species.[47,48,49,50,51] Indeed, their disordered multiple-element compositions lead to a strong possibility of SRO, *e.g.*, the preference for certain types of bonds within the first few neighbor shells. This is not particularly rare in conventional alloys[52,53] and glasses;[54] however, it could be argued that its existence would be even more likely in multiple principal element alloys[49,51,55] due to large number of elements and their equimolar concentrations. Recent DFT and MD simulations on the *fcc* CrCoNi alloy suggest that SRO can have a profound effect on critical parameters, notably the stacking-fault energy[55] and dislocation mobility;[56] accordingly, such local order could be an important factor in controlling mechanical properties.

In spite of extensive studies on the *bcc* transition metals, there are relatively few published studies of dislocation core structures, dislocation mobility, or the effect of chemical SRO for *bcc* RHEAs. Accordingly, the objective of the current paper is to employ DFT-based methods to compute the dislocation core structures in refractory HEAs and to explore the distribution of dislocation core energetics and its potential effect on Peierls barriers, focusing on the MoNbTaW system.

**RESULTS**

**Dislocation core structures in RHEAs**

To compute the core structures and Peierls potential for ½<111> screw dislocations in the refractory MoNbTaW HEA, we employ DFT calculations, making use of the Vienna *ab initio* simulation package;[57,58,59] details of the DFT calculations are provided in the Methods section. For screw dislocations in refractory HEAs, we employ a periodic supercell that contains 462 atoms, as illustrated in Figure 1a. The simulation cell contains a pair of dislocations with opposite Burgers vectors, in a nearly square quadrupolar arrangement[16] with triclinic symmetry to minimize any effects of periodic boundary conditions and image stress. This dipole approach was first introduced by Bigger *et al.*[60] and has been widely used in DFT calculations on dislocations.[16,17,20,61] The supercell adopted in current work was previously described by Weinberger[17] and Li *et al.*,[61] and was used to calculate dislocation core structures in pure *bcc* transition metals. In addition, since the size of supercell is fixed for all the simulations, the short periodic length might have some influence on the dislocation dipole energy due to its effect on the nature of the SRO. In our current model, we consider an equimolar MoNbTaW *bcc* RHEA,[37] and doubled the periodic length along the dislocation line direction of the original 231-atom model (see Methods section for further details) to minimize as much as possible correlations in the chemical order, as described in the following section.

The initial atomic configuration was generated by creating a special quasi-random structure (SQS) on the 462-atom supercell shown in Figure 1a. The SQS was generated using the Alloy Theoretic Automated Toolkit (ATAT) program.[62] The SQS methodology was used to minimize chemical correlations, and thus to provide a reference configuration corresponding to random substitutional disorder (*i.e.*, minimizing chemical SRO). This reference configuration was used in Monte-Carlo simulations to generate supercells with varying degrees of SRO, as described below. For each of the configurations with different level or SRO, we shifted the dislocation dipole over all the possible sites within the simulation cell, to statistically sample dislocation properties. The atomic positions in the system with the dislocation dipole were then relaxed to enable interrogation of the core structures and energies in different lattice sites within the RHEA supercell.

For each configuration representing a different degree of chemical SRO, we calculated 231 different structures with the dislocation dipole supercell, with the position of the cores initialized in different local environments. We find that the screw dislocations in *bcc* MoNbTaW HEAs maintain a compact core structure in most of the resulting relaxed structures, as illustrated by Figure 1b, which is similar to the case in pure *bcc* elements.[17,18] In a very few situations, the core can be extended on the (110) plane as shown in Figure 1c. The DFT calculations thus reveal the dominant role of compact cores for dislocations in the MoNbTaW alloy (see Supplementary Note 1 for further details).

**Local Chemical Short-Range Order in MoNbTaW RHEA**

Previously, a cluster expansion (CE) Hamiltonian in combination with Monte Carlo (MC) simulations have been developed to investigate the effects of SRO in MoNbTaVW and its quaternary sub-systems.[50] The ordering in the MoNbTaW RHEA alloy has been studied by Körmann *et al*.[63,64,65] This work revealed B2 long-range ordering at intermediate temperatures and phase decomposition in the ground state. For the present study, we employed a different approach (which nevertheless gives results in qualitative agreement with those of Kormann *et al*., as discussed below), chosen to enable the development of dislocation supercell models with representative degrees of chemical SRO. Our focus is specifically on the effect of SRO on the dislocation properties. For generating supercells with different degrees of SRO, similar to previous studies in *fcc* HEAs,[47,55] we applied a DFT-based lattice Monte Carlo (MC) approach to our 462-atom supercell model; details are described in the Methods section.

The supercell initiated with an SQS configuration was used as input for the MC simulations. The MC simulation samples swaps of atom types, following the Metropolis algorithm, and the entire simulation considers approximately 2100 such swaps, leading to the evolution of the energy shown in Figure 2a. Due to the limited number of MC steps and the lack of sampling of atomic displacements, the final configurations may differ from the true equilibrium state of SRO at the simulation temperature, although they appear to be quite close to the state of SRO as calculated by Kostiuchenko *et al.*[65] for high temperatures (~1200 K). However, the algorithm does lead to appreciable lowering of the energy, as shown in Figure 2a, and the pair-forming tendencies shown in Figure 2b are consistent with previous work on SRO in the same system using more comprehensive methods,[63,65] as discussed below. Thus, this method is used to generate representative samples with varying degrees of chemical SRO to explore the resulting effect on dislocation properties.

Similar to the conventional Warren–Cowley description[66] and the previous study for *fcc* HEAs,[55] we characterize the state of SRO using the so-called nonproportional number of local atomic pairs, $\Delta\delta_{ij}$, as described in more detail in the Methods section. Based on our calculations, the evolution of total potential energy and the overall chemical SRO ($\sum_{i,j}|\Delta\delta_{ij}|$) in the sample during the MC relaxations are plotted in Figure 2a. With respect to axes, the abscissa is the total potential energy change of the system and the ordinate is the overall chemical SRO of the system. As the MC simulation proceeds, the potential energy of system decreases monotonically while the chemical SRO increases at the same time. This clear trend indicates that chemical SRO is occurring in the system with the MC simulations. To quantify the effect of SRO on dislocations, three different samples from the simulation (*s1, s2, s3*) were chosen for further calculation of core structures and energies, indicated by the red arrows in Figure 2a. State *s1* represents the nearly random solid solution configuration with lowest magnitudes of the SRO parameters; *s2* represents an intermediate configuration with a medium level of SRO, and *s3* represents the configuration with the highest degree of SRO.

Figure 2b shows the quantitative values of $\Delta\delta_{ij}$ between all the species in the MoNbTaW alloy; the red dots show that the local SRO in state *s3* clearly deviates from the random solid solution. Preferred atomic pairings between Mo-Ta, Mo-Nb and Ta-W were observed as the $\Delta\delta_{ij}$ values are 0.308, 0.196 and 0.112, while unfavorable pairings between Mo-W and Ta-Nb were also apparent as the $\Delta\delta_{ij}$ values are -0.392 and -0.294. This result confirms the energetic preference for SRO in MoNbTaW alloys; moreover, the tendency to form SRO that we see here is consistent with previous studies using other methods[50,51,63,65] that have shown the Mo-Ta pairs are the most dominant contributors to the SRO, followed by Ta-W and Mo-Nb pairs.

**Distribution of dislocation core energies in *bcc* RHEAs**

After the introduction of SRO through MC relaxations, the dislocation dipole described in Figure 1a was created in samples *s1*, *s2* and *s3*. To sample over the distribution of local chemical environments for dislocation cores, the dislocation dipole was shifted over all the possible sites within the simulation cell leading to 231 different configurations for each of the three states of SRO. All the configurations with the dislocation dipole were then minimized, following the procedures described in the Methods section.

Figure 3 shows histograms of the supercell excess energies, *i.e.*, the energy difference between the supercell with and without the dislocation dipole, of all the configurations minimized at different SRO states. The histograms for the three SRO states are fit well by normal distributions (the fitted lines are also shown in Fig. 3). The green dash-dot line represents the energy distribution of the nearly random solid-solution sample *s1*. The blue dash line represents the sample *s2* with a medium degree of SRO and the red solid line represents the sample *s3* with highest degree of SRO. The mean values of the excess energies for the two samples with SRO differ by 0.38 eV (*s2*) and 0.87 eV (*s3*) from that for the most disordered sample (*s1*).

To compute dislocation core energies from these energies, we consider the components contributing to the supercell excess energy. The excess energy is the sum of the two dislocation core energies, the elastic energy arising from the dislocations, and a contribution from the diffuse antiphase boundary energy between the two cores created by the relative shift of the crystal by a Burger's vector across the planar "cut" region between the dislocations. This excess energy can thus be written as: $E = 2E^{core} + E_{elastic} + E_{DAPB}$. We note that in previous studies it has been shown the excess energies of the types of supercells used in this study can also be affected by the residual stress in the simulation box.[67,68,69] In Supplementary Figure 2, we plot the distribution of this residual stress on all the simulation cells, and the results rule out the correlation between the change in variance in the excess energies with these residual stresses. For what follows, we thus focus on the decomposition of the excess energies into core, elastic and DAPB contributions.

To first order, the elastic energy can be estimated using continuum theory as described by Clouet,[67,68,69] using the elastic constants and dislocation Burgers vector and a reasonable assumption for the core radius. For the simulation supercell used here, the DFT-calculated elastic contribution $E_{elastic}$ is estimated to be approximately 6.0 eV. Importantly for the analysis that follows, we find that the SRO has only an approximately 3% effect on the calculated elastic moduli (see further details in Supplementary Note 2), such that this local order is estimated to contribute only a 3% percent change (~0.17 eV between *s1* and *s3*) in the elastic energy contribution to the calculated excess energies. Details of the calculations of the elastic constants and elastic energy contribution in the dislocation dipole cell are provided in Supplementary Tables 1 and 2.

Another contribution to the average and variance in calculated excess energies for the supercells is associated with the cut plane between the two dislocation cores. When SRO is present, this cut plane leads to a contribution to the energy of the supercell arising due to the shift of adjacent planes, which disrupts the state of SRO and causes an excess energy $E_{DAPB}$. Following the convention in the literature, this planar defect is referred to as a diffuse antiphase-boundary (DAPB) and can be quantified through the so-called diffuse antiphase boundary energy per unit area ($\gamma_{DAPB}$). We have calculated the DAPB energy in our current system (see Supplementary Note 3), with the following results: for the state *s1* which represents the random solid solution, $\gamma_{DAPB}$ is ~3 mJ/m$^2$, *i.e.,* essentially zero within the accuracy of our statistical sampling. With increasing SRO, $\gamma_{DAPB}$ increases to 29 mJ/m$^2$ in state *s2* and 59 mJ/m$^2$ in state *s3* with the highest degree of SRO. $E_{DAPB}$ associated with the cut plane gives rise to an increasing contribution to the excess energy of the supercell: from 0.015 eV in state *s1* to 0.59 eV in state *s3*. Further, due to the important role of the variance in the core energy distribution, which will be discussed below, the variation in $E_{DAPB}$ due to the position of the cut plane as the locations of the dislocation cores are shifted as also calculated through DFT simulations. The standard deviation $\sigma_{E_{DAPB}}$ is approximately 0.15 eV for *s1,* 0.17 eV for *s2* and is increased to 0.26 eV for *s3*. Based on these data, we can further decouple the contribution of the variance in excess energies due to the two dislocation cores and the diffuse antiphase boundary. The details of these calculations are shown in Supplementary Note 3 and Supplementary Table 4.

Assuming that the excess energy shown in Figure 3 can be decomposed as $E = 2E^{core} + E_{elastic} + E_{DAPB}$, we can extract the distribution of core energies by subtraction of the contributions from elastic energy (see Supplementary Note 2) and the mean and variance of the DAPB energy (see Supplementary Note 3). Further details are given in Supplementary Note 4 and Supplementary Table 5. Figure 4 shows the average and variance values for the dislocation core energy in MoNbTaW for different SRO states. The average values are compared with the value in pure *bcc* transition metals from a previous DFT study.[18] The averaged core energy in MoNbTaW HEA is the highest compared with all its constituent pure elements. In addition, the SRO has only a marginal impact on the averaged dislocation core energy since it is a 1-D line defects; this result is in contract to the effect of SRO on 2-D planar defects, such as the stacking fault energy[55] or DAPB energies.

One important feature of Figure 3 is that the dislocation dipole energy follows a Gaussian distribution, which is an intrinsic feature of a HEA that differs from the pure element metals. Although the averaged core energy is not sensitive to SRO, the variance of the distribution is found to decrease with the increase of SRO. The standard deviation of the excess energy in Figure 3 for SRO state *s1* is 0.72 eV, which decreases to 0.36 eV in *s2* and to 0.37 eV in *s3*, *i.e.*, with lower degrees of SRO, the variance becomes more significant. The variances of dislocation core energies, decoupling the effect of the DAPB energy, are illustrated in Supplementary Table 4 and show similar trends. As described above, we conclude that the dominant contribution to the variance in supercell energy shown in Figure 3 arises from the variations in dislocation core energies; the results thus also demonstrate the role of SRO in changing the dislocation core energy distribution.

To illustrate the local spatial variation in core energies in the RHEA, and the effect of SRO on these variations, we plot 3D contours of the supercell excess energies and their 2D projection in the supercell in Figure 5. For simplicity, each dislocation dipole is treated as a single point located at the average spatial location of the two screw cores in the dipole; they are aligned in $[\bar{1}\bar{1}2]$ and $[1\bar{2}1]$ directions based on their relative positions. The excess energies normalized by the total length of the dislocation lines, which can be regarded as the depth of the Peierls valleys, are shifted to set the minimum value equal to zero. Based on these data, the left column of Figures 5a-c shows 3D contours of the Peierls valleys at different SRO states from *s1-s3* and the right column corresponds to their 2D projection. Note that this is not a minimum energy path (MEP) contour, since no transition-state data were included in these plots. For a pure element metal, the contour in Figure 5 would be that of a flat surface since the depth of the Peierls valley has a constant value. However, due to variations in local environment within the RHEA, the dislocation dipole energy in these alloys follows a normal distribution, as shown in Figure 3, which leads to rugged Peierls valleys contours, as shown in Figure 5. The maximum variation in the Peierls valleys is 0.9 eV/*b* in the near-random *s1* state; with increasing SRO, this decreases to 0.48 eV/b in *s2* and to 0.50 eV/b in *s3*. It is clearly visible in Figures 5a-c that the Peierls valley contours contains a rugged feature for the RHEA.

Similar to Figure 3, histograms of the differences in Peierls valley energy for different SRO states are shown in Figure 6 (see the Methods section for further details). As discussed further in the next section, the Peierls valley energy differences considered in Figure 3 are defined as $\Delta E = E_{d1} - E_{d2}$, where $E_{d2}$ is the excess energy of the supercell for one position of the dislocation dipole, and $E_{d1}$ is the excess energy when this dipole has shifted by glide to the neighboring Peierls valley in the $[\bar{1}\bar{1}2]$ direction. If we assume that the dislocation dipole energy follows the same Gaussian distribution shown in Figure 3, based on the properties of Gaussian distributions, the values of $\Delta E$ will also follow a Gaussian distribution but with a different variance: $\sim Normal(0, 2\sigma^2 - 2\sigma_{cov})$, where $\sigma_{cov}$ is the covariance of the excess energy for two neighboring positions of the dipole. In Figure 6, the average value of the energy difference is zero for all the three SRO states, as expected, and the variance of the fitted distribution from the DFT energy data agrees well with the prediction (details of the calculation of $\sigma_{cov}$ and $\sqrt{2}\sqrt{\sigma^2 - \sigma_{cov}}$ are given in Supplementary Note 5). Similarly, the values of Peierls valley energy difference (Δ*E*) defined above, corresponding to glide of the dislocations in the $[\bar{1}\bar{1}2]$ direction, are plotted in Figure 7 as a function of the initial position of the dipole, and are represented in both 3D contours and 2D projections. Standard analyses of transitions in complex systems are consistent with the basic trend that the energy difference between the final and initial states correlates with the change in the energy barrier. In the contours of valley energy differences shown in Figure 7, the values range from -0.30 to 0.30 eV/*b* in the *s1* state; these decrease to -0.12 to 0.15 eV/*b* in state *s2* and to -0.14 to 0.15 eV/*b* in the state *s3*. The fraction of these energies with relatively high values decreases with the increasing degree of SRO. These results, along with the change of distribution of dipole energies in Figure 3, demonstrate that the presence of SRO serves to narrow the distribution of dislocation core energies and decrease the spatial heterogeneity of dislocation core energies in the system. For reference, the Peierls barriers in the pure element constituent metals, Mo, Nb, Ta and W, calculated through the drag method, which is consistent with DFT study,[17] are also plotted on Figure 6. A significant amount of the Peierls valley energy difference (Δ*E*) during glide can be seen to have exceeded the highest value of the Peierls barriers in pure *bcc* elements. This rugged energy landscape and variance in core energies intrinsic in RHEA is anticipated to have a profound effect on the distribution of Peierls barriers, as explored further below.

**Peierls barriers of screw dislocations in *bcc* RHEAs with local chemical order**

In pure *bcc* transition metals,[17,20] the Peierls barriers for ½<111> screw dislocations can be computed from the energy pathway between two equilibrium samples, in which the dislocation dipole is uniformly translated along the $[\bar{1}\bar{1}2]$ direction on the {110} plane to the nearest neighboring site using the reaction coordinate method (also termed the "drag method"[70]) or the "nudged-elastic-band (NEB) method".[71] However, in a system with a complex energy surface, such as the RHEA considered here, the NEB method is computationally highly costly and difficult to converge. Alternatively, we have found that the "drag method" converges well.

Based on our tentative estimations of Peierls barriers through the "drag method", the most significant feature in the RHEA system is that the equilibrium energies of the dislocation dipoles are not constant due to the different local environments of the dislocation cores, compared with pure element metals. Thus, the potential energy of the initial configuration (where the reaction coordinate is 0), is generally not equal to that of the final configuration (reaction coordinate of 1). The shape of Peierls potential and the barrier values depend markedly on the relative energy difference between the initial and final configurations and can be divided into two distinct classes that we will refer to as Type-1 and Type-2 barriers, as shown in the schematic plot in the Figure 8a. Generally, the barrier value is higher than the potential energy difference between the final and initial configurations. When the potential energy difference between the initial and final configurations is small, or when the energy of the final configuration is smaller than that of the initial configuration, the barrier curves are usually Type-1, as shown by the red curve in Figure 8a. However, if the final configuration has a much higher potential energy than that of the initial configuration, the typical barrier curves under this condition will be like the blue curve shown in Figure 8a; these are referred to as Type-2 barriers, in which the Peierls barrier is dominated by the difference in potential energy between the initial and final configurations.

For pure element metals, we naturally expect 100% Type-1 shape barriers since the dislocation dipole energies are constant and the Peierls potential curve will be perfectly symmetric. For instance, based on our drag method calculations, the Peierls barriers in pure element metals, *i.e.*, Mo, Nb, Ta and W, range from 0.12 to 0.38 eV (0.03~0.09 eV/b if normalized by the total Burgers vector) in the current simulation geometry; they are plotted on Figure 6. If we take the highest Peierls barrier value in the pure *bcc* elements as the reference for the RHEA, it is found that when the core energies follow the Gaussian distribution, the rugged energy landscape and variance in RHEA will inevitably lead to another scenario during the calculation of the dislocation Peierls potentials, in which the final configuration has a much higher potential energy than that of the initial configuration, as shown by the right-side histogram in Figure 6, which is noted as a Type-2 barrier. Based on the histograms and contour of the Peierls valley energy difference shown in Figures 6-7, there is a significant degree of neighboring valley energy differences that have already exceeded the highest Peierls barriers found in pure *bcc* metals (0.37 eV or 0.09 eV/b in W). For the case of the random solid solution sample *s1,* as indicated by the green dash-dot line in Figure 3, which displays a relatively broad distribution of dislocation core energies, the probability of a Type-2 barrier will be higher. However, with progressively increasing SRO in samples *s2* and *s3*, the distribution of core energies narrows, as shown by the blue and red histograms in Figure 3. The lower variance of the core energies leads to fewer Type-2 barriers. In what follows, we argue that the variance or standard deviation of the core energies will lead to the asymmetric barriers and the variance itself is affected by the degree of SRO in the materials.

The transition from Type-1 to Type-2 barriers is highly dependent on the relative energy difference between the initial and final dislocation configurations. Here, we assume that for an alloy with a certain level of SRO, the dislocation dipole energy will follow a normal distribution:

$Normal(\mu, \sigma^2)$ (similar to Fig. 3), as shown in Figure 8b. If we assume that there are two random neighboring dipoles: dipole-1 and dipole-2, dipole-2 represents the initial configuration and will have a preference to glide to its final configuration dipole-1. The energy of these two dipoles are written as $E_{d1}$ and $E_{d2}$. For the transition from a Type-1 to a Type-2 barrier, we postulate that there exists a critical energy difference $E_{critical}$ that when $E_{d1} - E_{d2} > E_{critical}$, the Peierls barrier will become a Type-2. Based on our assumptions for the distribution in dipole energies in Figure 8b, the energy of dipole-1 and dipole-2 are: $E_{d1} \sim Normal(\mu, \sigma^2)$, $E_{d2} \sim Normal(\mu, \sigma^2)$. The energy difference between the two dipoles is then $E_{d1} - E_{d2} \sim Normal(0, 2\sigma^2 - 2\sigma_{cov})$. Since the energies of two neighboring dislocation dipoles are not independent, we need to consider the covariance $\sigma_{cov}$ between these $E_{d1}$ and $E_{d2}$ values (see details in Supplementary Note 5). Thus, the probability of observing a Type-2 barrier for this condition can be written as:

$$P_{type2} = P(E_{d1} - E_{d2} > E_{critical}) = 1 - P(E_{d1} - E_{d2} \leq E_{critical}) = 1 - \Phi(\frac{E_{critical}}{\sqrt{2}\sqrt{\sigma^2 - \sigma_{cov}}})$$
(1)

where $\Phi$ is the standard Normal cumulative distribution function and $\sigma_{cov}$ is the covariance between the energy of two neighboring dislocation dipoles.

Based on this equation, the probability of a Type-2 barrier is a function of $E_{critical}$, the standard deviation $\sigma$ (or variance) of the dipole energy distribution and covariance between energy of two neighboring dislocation dipoles. In Figure 8c, we plot $P_{type2}$ as a function of $\sigma$ for two different values of $E_{critical}$ with $\sigma_{cov} \in [-0.8\sigma^2, 0.8\sigma^2]$. The two $E_{critical}$ values were chosen as 0.4 eV and 0.6 eV, which is slightly higher than the Peierls barrier calculated in W (0.37 eV). These curves clearly demonstrate that the probability of a Type-2 barrier will increase monotonically with the standard deviation $\sigma$, which is also correlated with the state of SRO. For a single screw dislocation, rather than the dislocation dipole geometry considered in this study, we can obtain similar results as $P_{type2} = 1 - \Phi(\frac{e_{critical}}{\sqrt{\sigma^2 - \sigma_{cov}}})$. This is discussed in further detail in Supplementary Note 6. This analysis highlights the origin of the Type-2 Peierls barrier and its correlation with the state of SRO in the RHEAs. Although we cannot obtain accurate Peierls barriers in the current study based on DFT calculations alone, we can conclude that in simple terms, the unique variance of dislocation core energies in RHEA, which is also influenced by the SRO, enhances the probability of observing Peierls barriers of Type-2, which will finally influence the dislocation morphologies and their motion.

**DISCUSSION**

Using first principles calculations of dislocation energies and the differences in Peierls valley energies in *bcc* RHEAs, our results reveal fundamental differences between behavior in the multiple principal element alloys and a pure metal or dilute solution. The variation in local chemical environments within the RHEAs lead to a distribution of dislocation core energies for different dislocation segments; moreover, the characteristics of this core energy distribution are significantly influenced by the presence of SRO. In contrast, all the local environments are constant in pure metals and would be expected to show much smaller distributions for dilute solutions.

With our present DFT calculations, although we have doubled the thickness of the sample, the dimension of the out-of-plane direction is still limited to only two Burgers vectors. The calculated core energies and Peierls potentials thus represent the local characteristics of a small straight segment of dislocation line. When considering a long dislocation line gliding in the RHEA, due to the Gaussian distribution of local energies of dislocation segments, described in Figure 3, the dislocation line will prefer to form a wavy shape to reduce the total potential energy. For alloys with multiple principal elements in equal molar ratios, statistically the composition fluctuation always exists even for a random solid solution.

The Peierls potential plays a crucial role in governing dislocation motion. Here, we have identified two types of Peierls barriers in the *bcc* RHEA which depend critically on the energy distribution of the dislocation segments. Considering a long dislocation motion associated with kink-pair theory,[72] it is extremely difficult for some segments gliding through the path of the Type-2 barriers due to its high magnitude. Under such circumstances, these segments can become pinned or are forced to glide on alternative planes or in different directions. This will serve to facilitate cross slip, dislocation multiplication and the formation of wavy dislocation lines, all of which will eventually enhance the strength and ductility of the material at the macroscale due to homogenization of plastic strains.[73] Indeed, such a form of wavy slip and enhanced mechanical properties has been reported for a *bcc* TiZrNbHf RHEA with short-range ordered (O,Ti,Zr)-complexes.[73] Recently, a theory[74] developed for screw dislocation strengthening in RHEAs has been presented based on the assumption that screw dislocations will naturally adopt a kinked configuration. Along with the MD simulations of the NbTaV alloy[74], our DFT data, as shown in Figures 3-5, strongly supports the idea that dislocation lines in this and related RHEAs would tend to form a kinked structure.

In summary, we have systematically studied the dislocation core energy, diffuse antiphase boundary energy, dislocation dipole energy distribution and Peierls valley energy differences in a *bcc* MoNbTaW refractory high-entropy alloy using DFT calculations, considering the effects of chemical SRO. Similar to the pure *bcc* transition metals, compact cores were found to dominate in screw dislocations in the *bcc* MoNbTaW RHEA. The average core energy of a screw dislocation is higher in the current RHEA compared with the pure *bcc* transition metals; however, SRO is found to have only a negligible effect on the average equilibrium energy of this line defect. However, the DAPB energy is found to correlate strongly with the SRO state that could potentially influence the dislocation mobility. In addition, the dislocation core energies were found to follow a Gaussian distribution with the increasing degree of SRO resulting in a progressively lower variance of the distribution of core energies. Resulting from the intrinsic fluctuation of core energies in HEAs, two types of Peierls barriers were discovered, which depend on the difference in core energies between initial and final configurations. By comparison with pure *bc*c transition metals, the Peierls barrier of screw dislocations in *bcc* RHEAs is expected to be higher due to the formation of Type-2 Peierls barriers from the variance of core energy distribution. The findings from the present work highlight the effect of the variance in core energy distributions in influencing dislocation Peierls potentials and suggest important consequences on dislocation morphology and activity, which is an intrinsic feature of HEAs. As these characteristics are heavily influenced by SRO, such local ordering may have a significant impact on the mechanical properties of refractory high-entropy alloys.

## METHODS

**Lattice constant determination and simulation cell with dislocation dipole**

The lattice constant of the equimolar MoNbTaW HEA was determined by relaxing the 64-atom quaternary quasi-random structure (SQS)[75] provided by Gao *et al.*[76] The calculated lattice constant was 3.230Å and was adopted in all simulations. For the simulation cell with dislocation dipole, we first defined $\boldsymbol{e}_1 = a_0[\bar{1}\bar{1}2], \boldsymbol{e}_2 = a_0[1\bar{1}0], \boldsymbol{e}_3 = a_0/2\,[111]$. Then, the supercell with a dislocation dipole was built with three edges, $\boldsymbol{h}_1 = 7\boldsymbol{e}_1, \boldsymbol{h}_2 = 3.5\boldsymbol{e}_1 + 5.5\boldsymbol{e}_2 + 0.5\boldsymbol{e}_3, \boldsymbol{h}_3 = 2\boldsymbol{e}_3$, to contain 462 atoms. The periodic length along the dislocation line direction, $\boldsymbol{h}_3$, was twice the magnitude of the Burgers vector.

**DFT-based Monte Carlo simulations**

Monte Carlo (MC) simulations were performed using the supercell geometry described above. For the initial condition in these simulations, the sample was generated as an SQS model of the random alloys. The temperature employed in the MC simulations was 500 K. Energy calculations were performed using the Projector Augmented Wave (PAW) method,[77,78] as implemented in the Vienna *ab initio* simulation package.[57,58,59] A plane wave cut-off energy of 400 eV was employed, and the Brillouin zone integrations were performed using Monkhorst–Pack meshes[79] with a 3 ×1 × 1 grid, where the first index corresponds to the direction along the dislocation line. Projector augmented wave potentials[78] were employed with the Perdew–Burke–Ernzerhof generalized-gradient approximation for the exchange-correlation function.[80] Lattice MC simulations were then conducted similar to the methods utilized by Tamm *et al.*[47] and Ding *et al.*[55], which included swaps of atom types with the acceptance probability based on the Metropolis–Hastings algorithm.[81] In the current MC simulations, a total of 2094 swaps were conducted and 471 swaps were accepted. For the choice of PAW potentials, 6 valence electrons were used for Mo and W, 5 valence electrons for Ta, and 11 valence electrons for Nb.

**Core structure and Peierls valley energy differences**

Following the MC simulations, the dislocation dipole was introduced into the sample at all possible sites. All configurations with the dislocation dipole were then relaxed through a conjugate-gradient algorithm using VASP with the settings described above, but with a denser k-point mesh of 7 ×1 ×1. Atomic positions were relaxed with a convergence criterion on forces of $10^{-2}$ eV/Å. For each relaxed sample selected as the initial configuration, we chose the sample with a nearest dislocation dipole on the same {110} plane and displaced in the $[\bar{1}\bar{1}2]$ direction as the final configuration to calculate the valley energy differences between these two neighboring dipoles. Further details can be found in Supplementary Figure 3.

**Local chemical short-range order parameter**

Similar to the definition described by Ding *et al.*[55], which was modified from the Warren–Cowley parameter,[66] we defined the nonproportional number of local atomic pairs, $\Delta\delta_{ij}$, to quantify the chemical ordering around an atomic species for the combined first and second nearest-neighbor shells in the *bcc* structure, for which the corresponding coordination numbers are $N = 14$. The value of $\Delta\delta_{ij}$ was then calculated as:

$$\Delta\delta_{ij} = N(p_{ij} - p_{ij}^{ideal}) ,$$

(2)

where $N = 14$ is the coordination number of first and second nearest-neighbor shells in the *bcc* structure, $p_{ij}$ is the actual probability of bonds between atoms of type *j* and type *i* in the sample, $p_{ij}^{ideal}$ is the ideal probability of bonds between atoms of type *j* and type *i* for the random solid solution case based on the species concentrations. $\Delta\delta_{ij}= 0$ for the case of a random solution. The overall SRO is represented by the sum of all the $|\Delta\delta_{ij}|$ for all species ($SRO = \sum_{i,j}|\Delta\delta_{ij}|$).

## DATA AVAILABILITY

The data that support the findings of this study are available from Dr. Sheng Yin (email: shengyin@berkeley.edu) upon reasonable request.

## CODE AVAILABILITY

The DFT calculations were performed with the Vienna *ab initio* simulation package. All the other codes that support the findings of this study are available from Dr. Sheng Yin (email: shengyin@berkeley.edu) upon reasonable request.

## ACKNOWLEDGEMENTS

This work was supported by the US Department of Energy, Office of Science, Office of Basic Energy Sciences, Materials Sciences and Engineering Division, under contract no. DE-AC02-05-CH11231 within the Damage-Tolerance in Structural Materials (KC 13) program. The study made use of resources of the National Energy Research Scientific Computing Center (NERSC), which is also supported by the Office of Basic Energy Sciences of the U.S. Department of Energy under the same contract number.

## COMPETING INTERESTS

The authors declare no competing interests, financial or otherwise.

## AUTHOR CONTRIBUTIONS

S.Y., M.A. and R.O.R. conceived the research, S.Y. and J.D. performed the numerical simulations and analyzed the results, and M.A. and R.O.R. supervised the research. All authors contributed to the writing of the manuscript.

**LIST OF FIGURES**

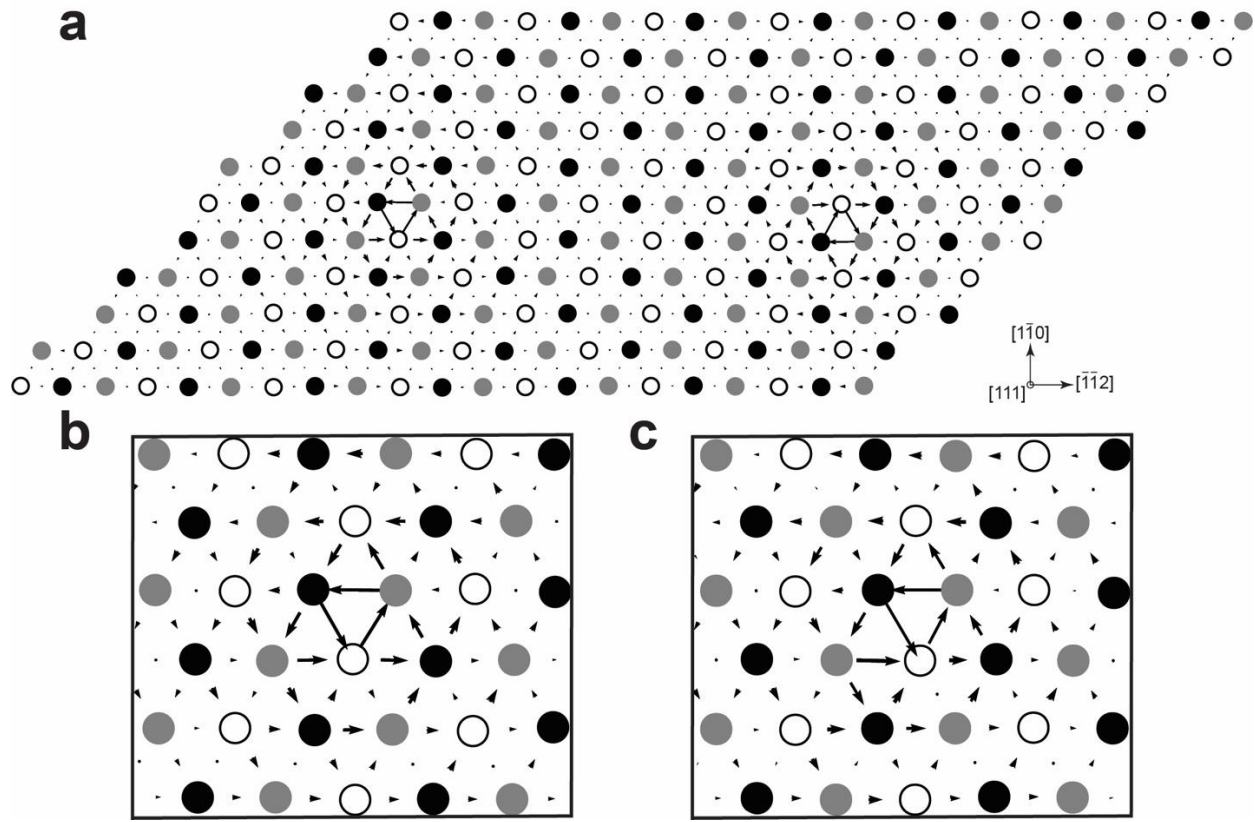

**Fig. 1 | Dislocation dipole model and structure of dislocation cores in an equimolar MoNbTaW *bcc* RHEA. a,** Differential displacement map of the dislocation dipole model. **b,** Close-up view of the structure of a compact core. **c,** Close-up view of the structure of a slightly non-compact core. The colors represent the relative position of atom in the [111] direction.

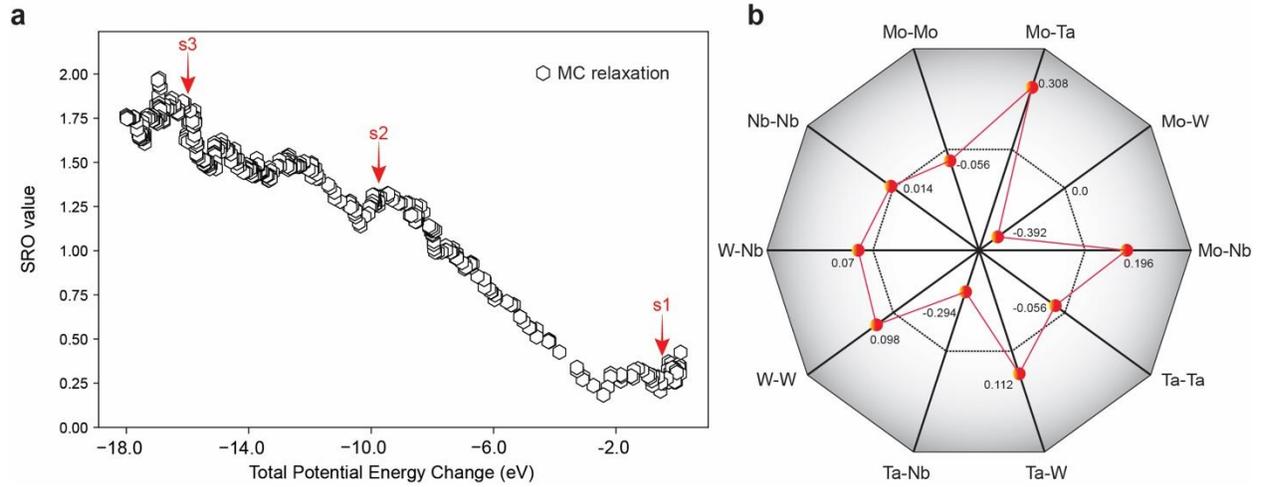

**Fig. 2 | Evolution of energy and local chemical SRO in the MoNbTaW RHEA. a,** Potential energy change *vs*. SRO parameter during the MC relaxation. Three states (*s1, s2, s3*) with different levels of SRO as indicated by red arrows were chosen for calculations of the dislocation cores and Peierls potentials. **b,** The detailed values of $\Delta\delta_{ij}$ for all atom pairs. The red lines and dots represent state *s3* with SRO and the dashed lines represent the ideal random solid solution case.

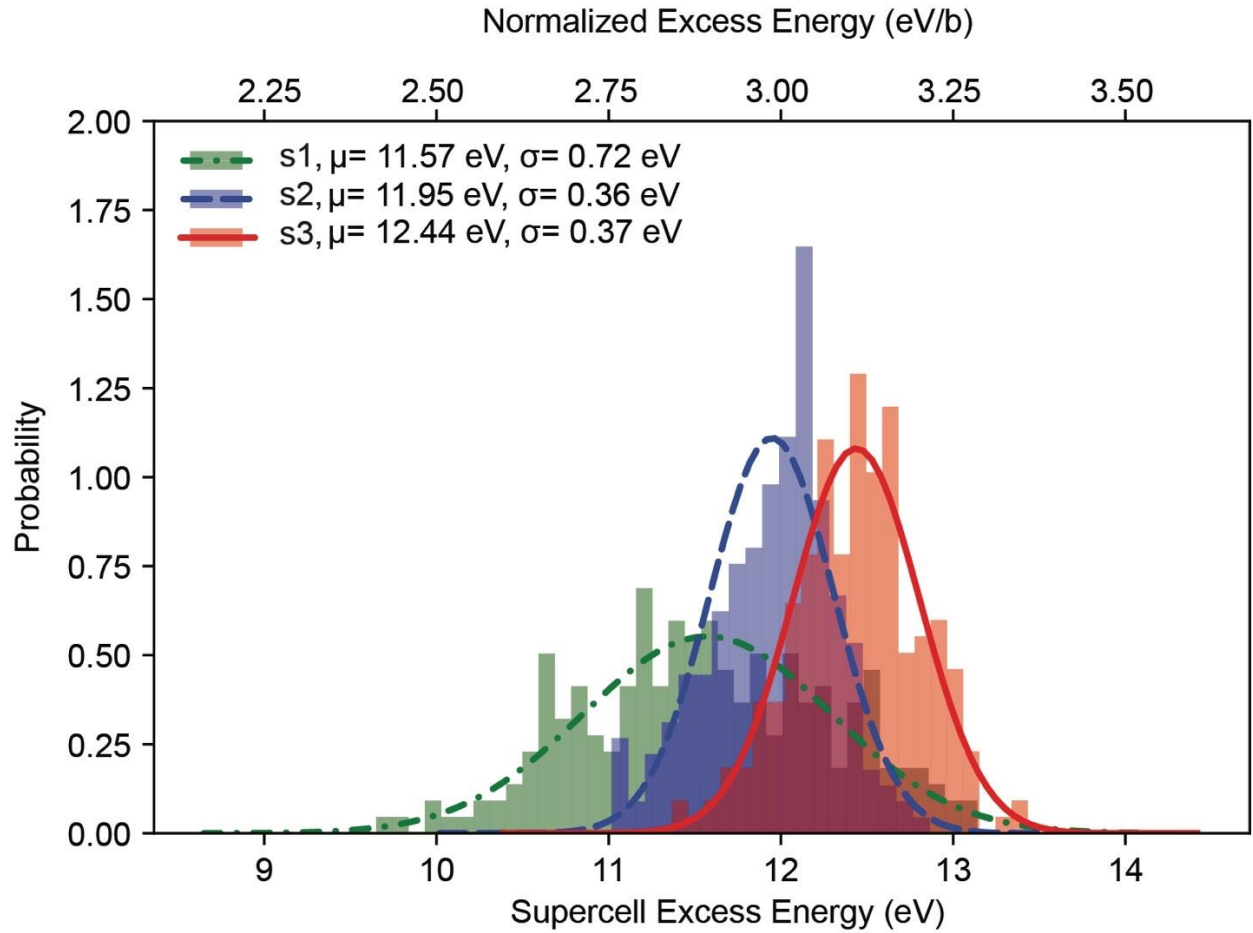

**Fig. 3 | Histograms of supercell excess energies for different levels of SRO in MoNbTaW.** Histograms of supercell excess energies for varying positions of the dislocation cores for three different states (*s1, s2, s3*) of SRO, fitted with a Gaussian distribution. The normalized excess energy indicated on the upper *x*-axis scale corresponds to the supercell excess energy divided by the total Burgers vector length in the supercell (*i.e.*, 4*b*). Note that state *s1* represents the random solid-solution state with minimum SRO, *s2* has a medium level of SRO, and state *s3* has the highest level of SRO. Mean and variance values for each of the Gaussian fits for different states of SRO are indicated in the upper legend.

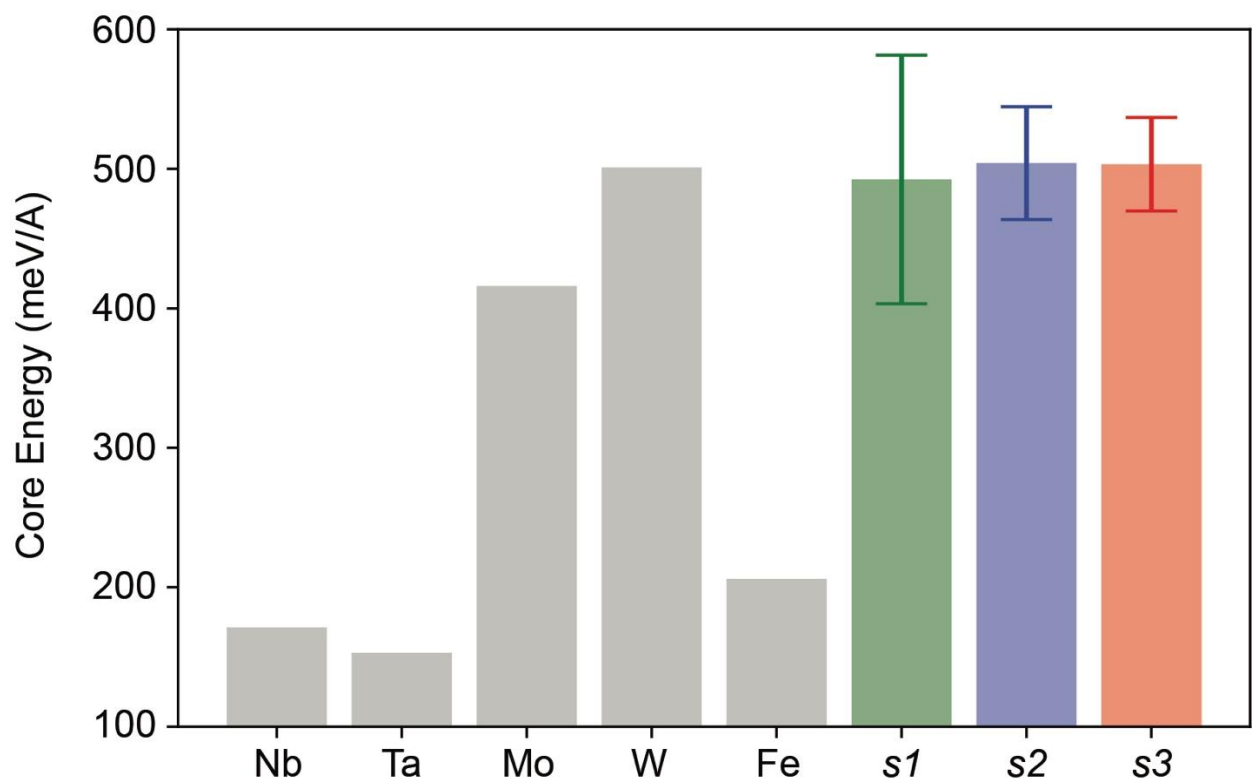

**Fig. 4 | Screw dislocation core energies in MoNbTaW, compared with the constituent pure *bcc* transition metals.** The labels *s1*, *s2* and *s3* are current results for MoNbTaW, with *s1* corresponding to minimum SRO, *s2* a medium level of SRO, and *s3* the highest degree of SRO. Core energy data for pure *bcc* transition metals are reproduced from a previous DFT study.[18] Error bars in the results for MoNbTaW correspond to standard deviations in the values derived by sampling different local environments. The core cutoff radius $r_c = 3.0$ Å in all cases. (See detail in Supplementary Note 4)

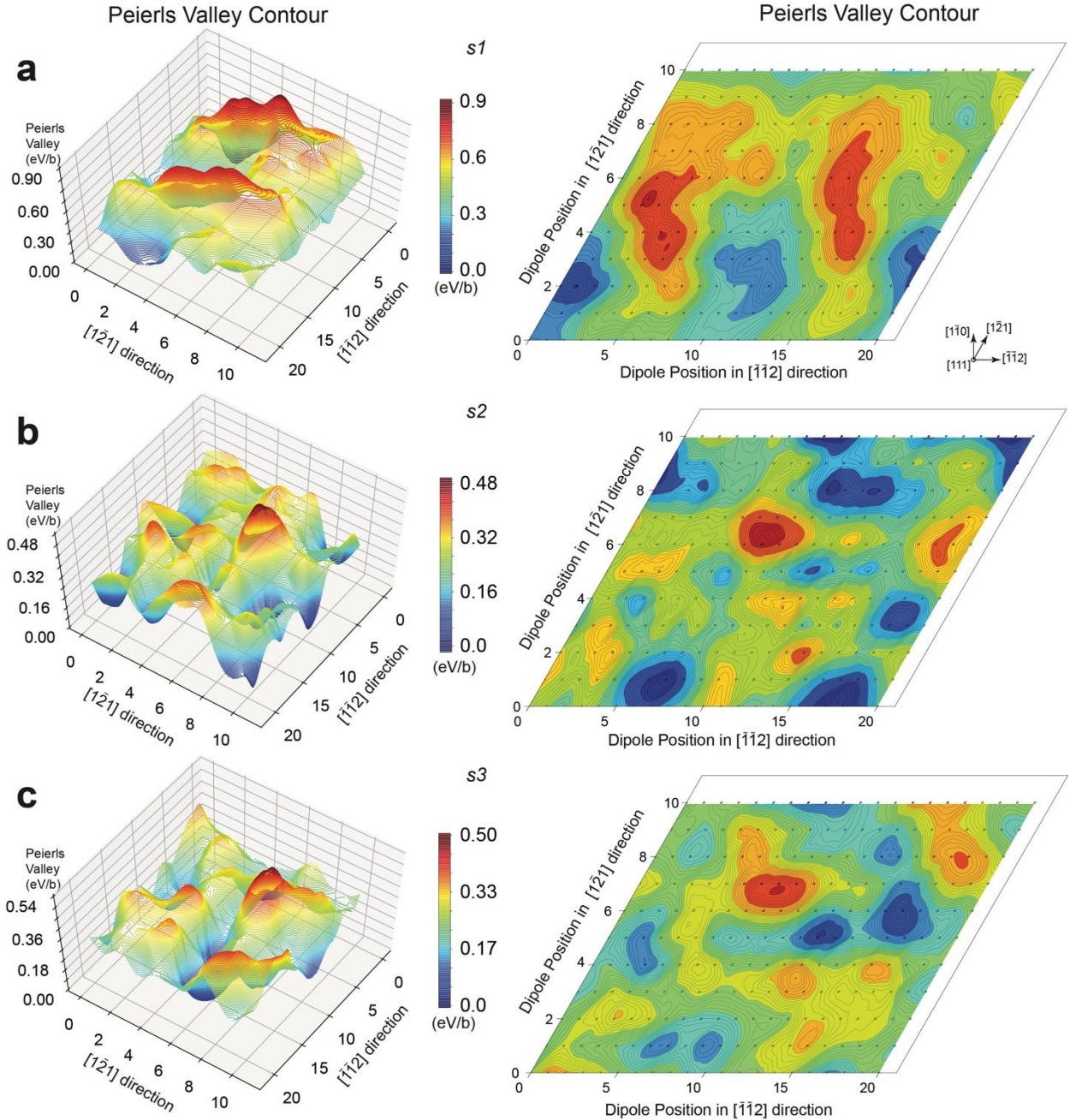

**Fig. 5 | Contours of Peierls valleys at three levels of SRO in the MoNbTaW supercell (no data for transition states is included). a,** random solid solution state *s1* with minimum SRO. **b,** state *s2* with a medium level of SRO. **c,** state *s3* with the highest level of SRO. The relative positions of the dislocation dipole are projected on (111) plane and aligned in $[\bar{1}\bar{1}2]$ and $[1\bar{2}1]$ directions, as shown by the black dots on the right column figures. The left column contains the 3D contours and the right column shows the corresponding 2D projection. The contours were plotted by interpolating data points on grids through bivariate spline.

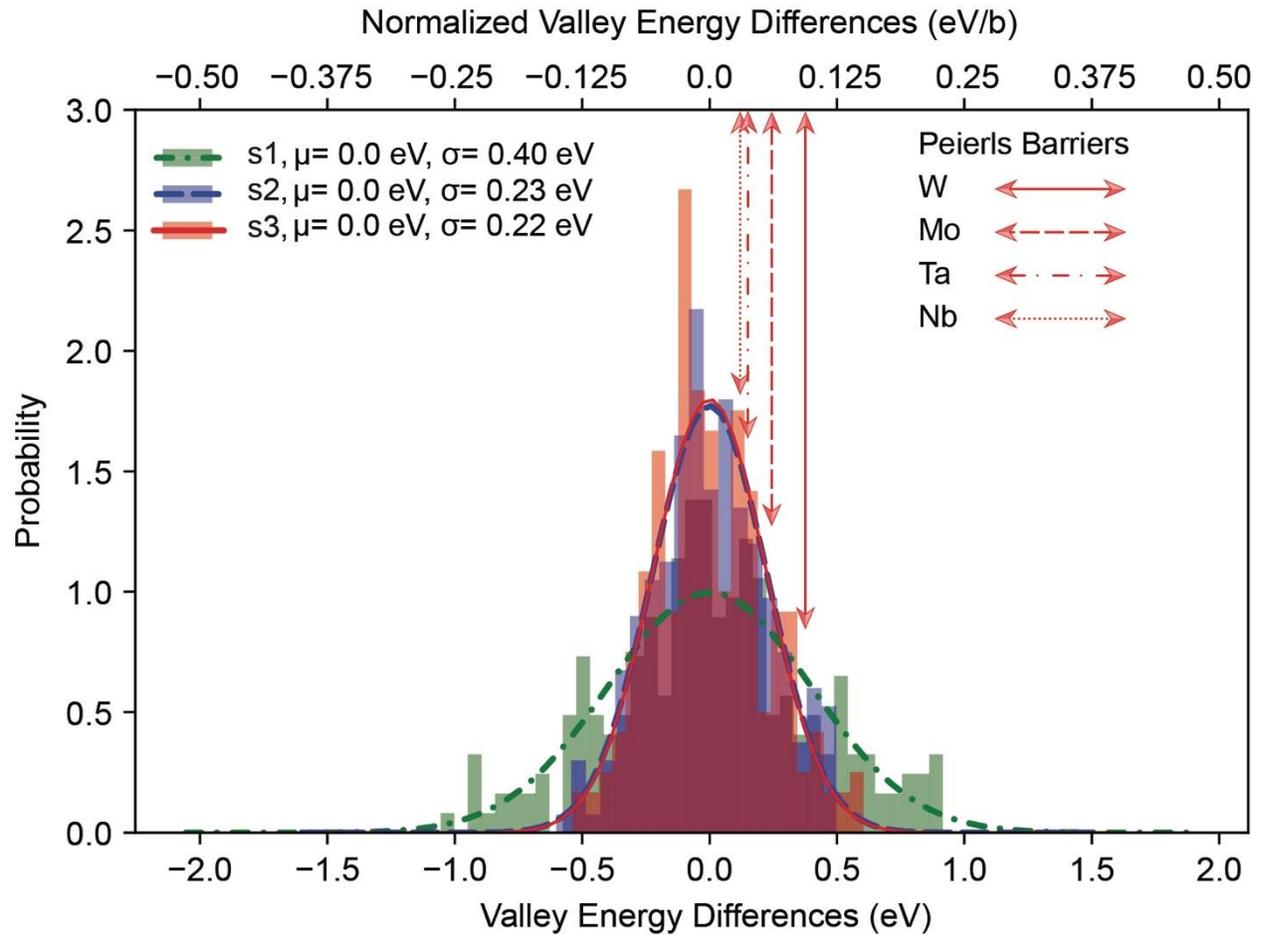

**Fig. 6 | Histograms of Peierls valley energy differences for different levels of SRO in MoNbTaW.** Histograms of the differences in supercell energies for dislocation cores in neighboring sites, sampled over the different local environments for three different states (*s1, s2, s3*) of SRO. Each of the histograms is fit with a Gaussian distribution, with associated fitted mean and variance values given in the upper legend. The normalized valley energy difference in the upper legend is the supercell dipole energy difference for neighboring sites, divided by the total Burgers vector length in the supercell (*i.e.*, 4*b*). The Peierls barriers of pure *bcc* metals calculated through drag method are also plotted for reference, reproduced from previous DFT study.[17]

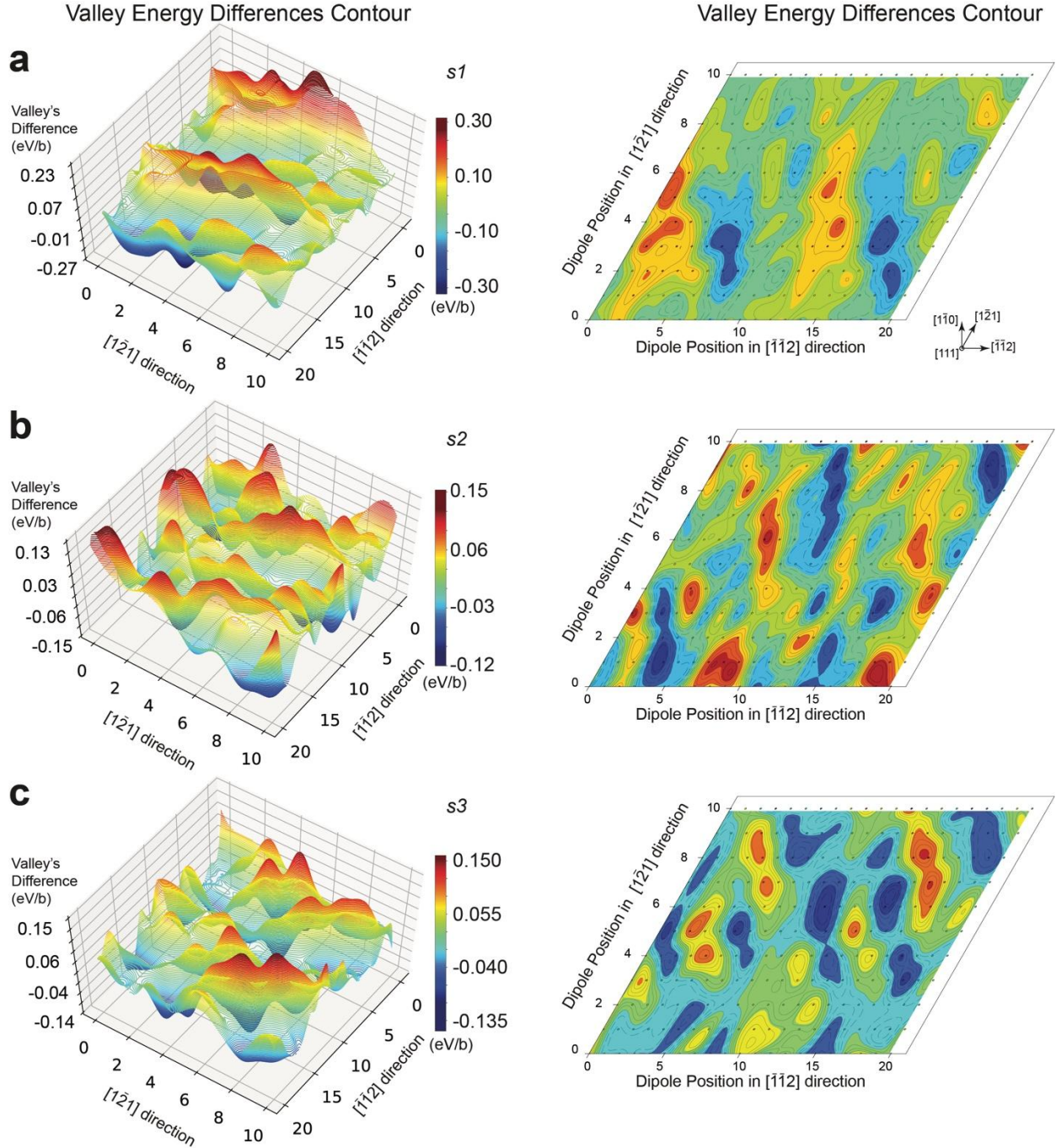

**Fig. 7 | Contours of the difference in the Peierls valley energy at three levels of SRO in the MoNbTaW supercell. a,** random solid solution state *s1* with minimum SRO. **b,** state *s2* with a medium level of SRO. **c,** state *s3* with the highest. The relative positions of the dislocation dipole were projected on (111) plane and aligned in $[\bar{1}\bar{1}2]$ and $[1\bar{2}1]$ directions as shown by the black dots on the right column figures. The glide direction of the dislocation dipole is along the $[\bar{1}\bar{1}2]$ direction (see Methods), and the plotted energy difference corresponds to the difference in energy between the final state (after glide) and

initial state for each position of the dislocation cores. The left column contains the 3D contours and right column is the corresponding 2D projection. The contours were plotted by interpolating data points on grids through bivariate spline.

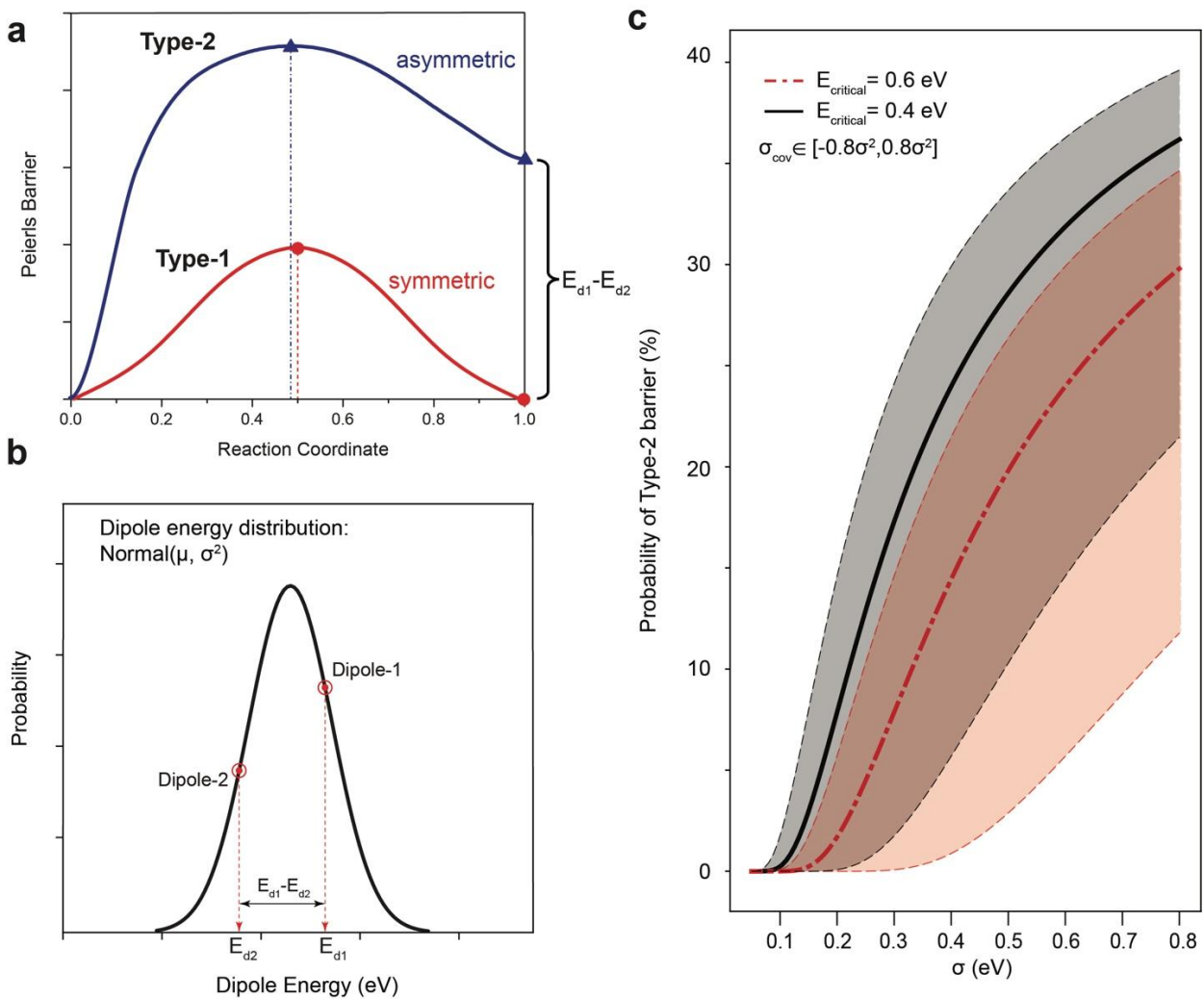

**Fig. 8 | Two types of barriers: Symmetric and asymmetric Peierls barrier curves in the RHEA. a,** Schematic figure of two types of Peierls barrier obtained in the MoNbTaW RHEA. **b,** Schematic figure of the distribution of dipole energy and two random neighboring dipoles. **c,** Probability of a Type-2 barrier as a function of the standard deviation of the dipole energy.

# Supporting Information

## *Ab initio* modeling of the energy landscape for screw dislocations in body-centered cubic high-entropy alloys

**Sheng Yin**[1,2], **Jun Ding**[2], **Mark Asta**[1,2*] and **Robert O. Ritchie**[1,2*]

[1]Department of Materials Science & Engineering, University of California, Berkeley, CA 94720, USA

[2]Materials Sciences Division, Lawrence Berkeley National Laboratory, Berkeley, CA 94720, USA

*To whom correspondence may be addressed; email: mdasta@berkeley.edu or roritchie@lbl.gov

**Supplementary Note 1: Compact *vs*. non-contact dislocation cores**

In the compact cores shown in Supplementary Figure 1, the differential displacement map is similar to the displacement of an ideal screw dislocation obtained by the Volterra construction, with the red arrow and blue arrow forming an equilateral triangle. Around the dislocation cores, the length of red arrows is written as $l_{ri}$ and the length of blue arrows is written as $l_{bi}$, where $i = $ 1,2,3. Ideally, in a perfect Volterra construction, $l_{ri} = constant1$, $l_{bi} = constant2$ for $i = $ 1,2,3 and $l_{ri} = 2l_{bi}$. In RHEAs, due to the complexity of numerous chemical species, local atomic environments and lattice distortions, $l_{ri}$ and $l_{bi}$ are not expected to be constant. Here, we define the ratio $r = \min(l_{ri})/\max(l_{bi})$, according to the observations that if $r < 1.10$, the dislocation core can be classified as a non-compact core. Supplementary Figure 1 shows two more examples of non-compact cores. However, among all the dislocation cores examined, compact cores are the most dominant, with the fraction of non-compact cores (based on the criterion for $r$ given above) being around ~1%.

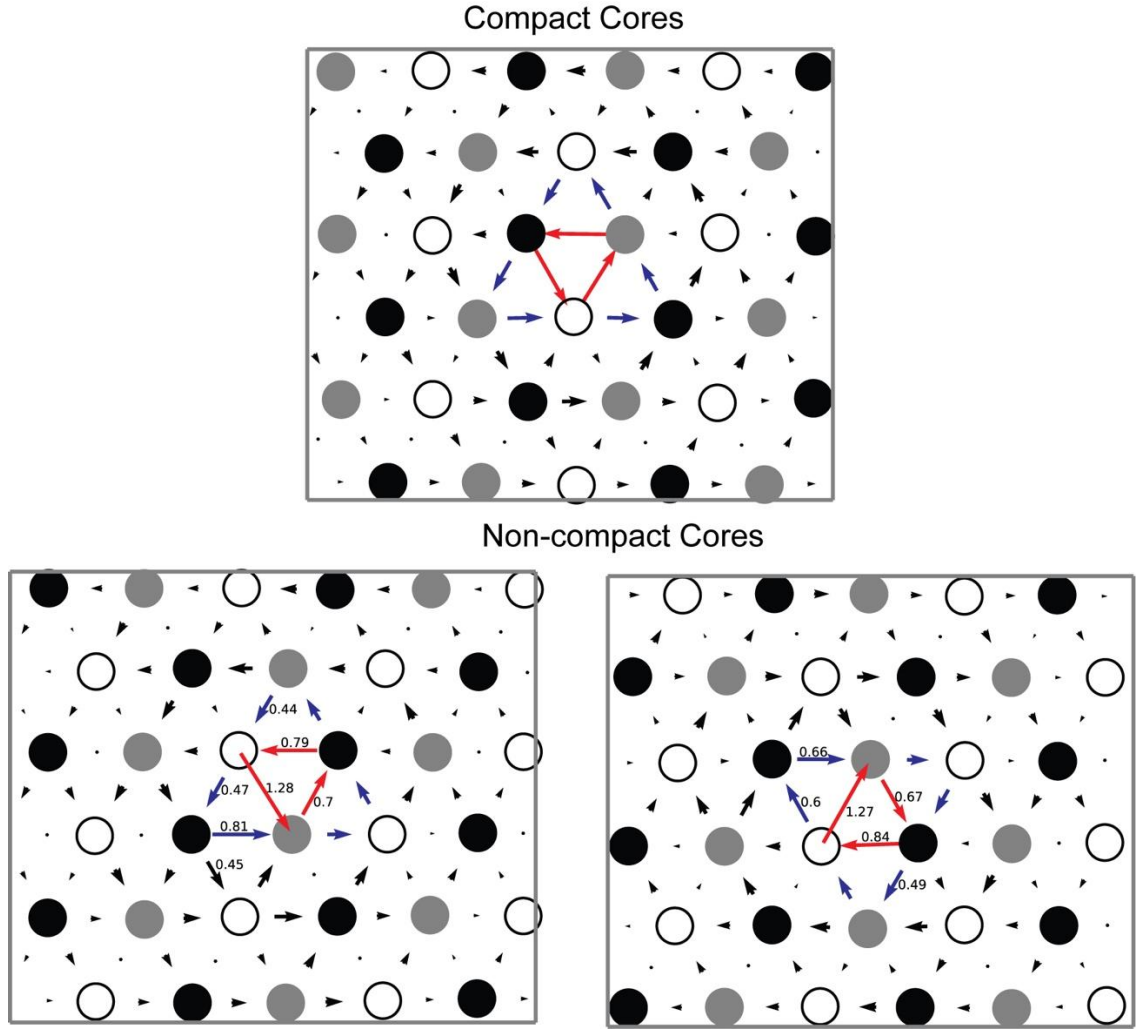

**Supplementary Figure 1 | Definition of compact cores and non-compact cores in differential displacement maps.** The upper panel represents a typical compact core, and the lower two are examples of non-compact cores, as described further in Supplementary Note 1.

## Supplementary Note 2: Contribution from elastic energy of dislocation dipole

*Elastic Constant Calculation:* The calculation of the elastic constants was performed using the relaxed structures without dislocation dipoles in three different SRO states: *s1, s2* and *s3*, as obtained from the Monte-Carlo simulations. Similar to the method described by de Jong *et al.,*[1] 24 unique deformation mappings are constructed for each sample, corresponding to six independent deformation modes, which include uniaxial deformation in three axial directions ($\varepsilon_{11}, \varepsilon_{22}, \varepsilon_{33}$) and simple shear deformation in three different directions ($\varepsilon_{12}, \varepsilon_{23}, \varepsilon_{31}$). To calculate the elastic constants, four values for the strain ($-2\varepsilon_0, -\varepsilon_0, \varepsilon_0, 2\varepsilon_0$) were applied for each of the six deformation modes. The value of $\varepsilon_0 = 0.005$ was chosen for uniaxial deformation and $\varepsilon_0 = 0.002$ was chosen for simple shear deformation. The components of the Cartesian stress tensor are calculated from the VASP runs, allowing ionic relaxations for each state of imposed homogeneous strain. By assuming cubic symmetry, all components of the elastic tensor can be determined by fitting the calculated stresses to the applied Green-Lagrange strain. The three independent elastic constants were then calculated by averaging all the calculated stress states. Supplementary Table 1 shows the averaged elastic moduli of MoNbTaW in three SRO states: *s1, s2* and *s3*. In Körmann *et al.*'s work,[2] the bulk modulus for the A2 random solution MoNbTaW alloy is 231 GPa and is only increased by 2 GP to 233 GPa for the B2(MoW;NbTa) structure. For the current calculations in Supplementary Table 1, the bulk modulus of the random solution MoNbTaW (s1) is 239 GPa and increases to 240 GPa for *s3*. The results show that the SRO has marginal impact (at most a few percent) on the elastic moduli of the alloys, which is also consistent with previous calculations of the change in bulk modulus due to ordering in HEAs.[2]

As discussed in the main text, the excess supercell dipole energy shown in Figure 3, *i.e.*, the energy difference between the supercell with and without the dislocation dipole, is the sum of the two dislocation core energies, the elastic energy, and a contribution from the diffuse antiphase boundary energy between the two cores; this can be written as: $E = 2E^{core} + E_{elastic} + E_{DAPB}$. Following the work of Clouet *et al.*,[3] we have calculated the total elastic energy ($E_{elastic}$) in the cell based on the averaged elastic moduli shown in Supplementary Table 1 and the supercell configuration. We neglected the local environment of the HEA and approximated the calculation of $E_{elastic}$ as the same as the pure element cell. In this calculation, the core cutoff radius was equal to 3.0 Å and $E_{elastic}$ contains the elastic energy of the dipole contained in the supercell and its elastic interaction with periodic images (40 periodic images in $[\bar{1}\bar{1}2]$ and $[1\bar{1}0]$ were considered here). The value of $E_{elastic}$ for *s1* is 6.05 eV and increases merely 0.17 eV due to the effect of SRO to reach 6.22 eV in *s3* (shown in Supplementary Table 2). Since the impact of SRO on elastic moduli is small, it also has marginal impact on the $E_{elastic}$ in the simulation cell. The contribution of $E_{DAPB}$ will be discussed in Supplementary Note 3.

**Supplementary Table 1 | Elastic modulus from DFT calculation.** Averaged elastic modulus of MoNbTaW in three SRO states. Units in GPa.

| Sample | $C_{11}$ | $C_{12}$ | $C_{44}$ | Bulk Modulus (Voigt average) | Shear Modulus (Voigt average) |
|---|---|---|---|---|---|
| s1 | 336.2 | 190.6 | 89.9 | 239.1 | 83.0 |
| s2 | 335.1 | 191.3 | 89.5 | 239.3 | 82.5 |
| s3 | 338.8 | 190.1 | 92.3 | 239.7 | 85.2 |

**Supplementary Table 2 | Elastic energy in the dislocation dipole cell.** Units in eV.

| Sample | $E_{elastic}$ |
|---|---|
| s1 | 6.05 |
| s2 | 6.03 |
| s3 | 6.22 |

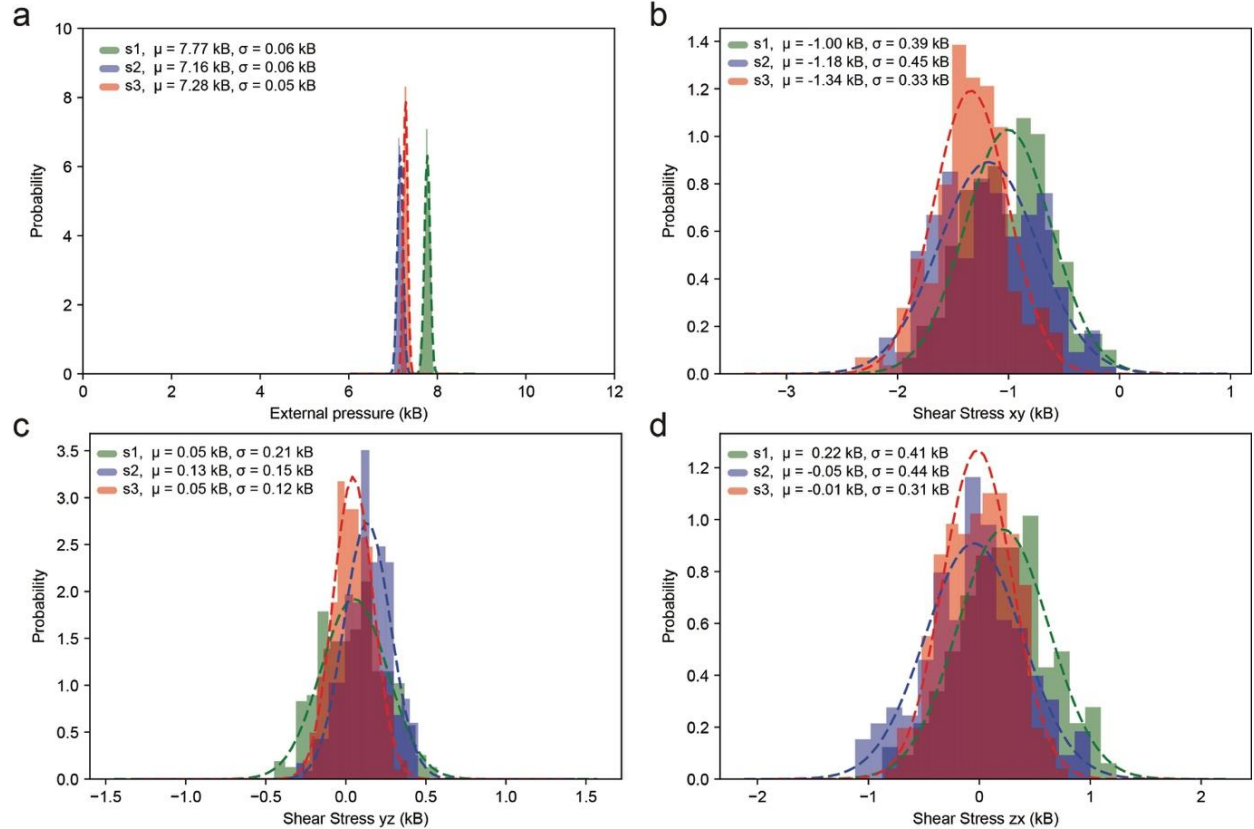

**Supplementary Figure 2 | Distribution of pressure and shear stresses on the simulation cells.** Histograms of residual stress in the simulation cells with dislocation cores in different sites in the samples with three different states of SRO, each fit with a Gaussian distribution. **a**, Histogram of the pressure (from the trace of the stress tensor), **b-d,** Histograms of residual shear stress. The green dash-dot line represents the nearly random solid solution state *s1* with minimum SRO. The blue dash line represents the state *s2* with a medium level of SRO. The red solid line represents the state *s3* with the highest level of SRO. The standard deviation of the residual pressure shows small changes in magnitude going from *s1* to *s3*. The magnitude of the residual shear stresses is much smaller than the residual pressure; the standard deviations do not show monotonic relationships with the degree of SRO. If we assume the bulk modulus $K\sim250$ GPa, the strain energy due to the current external pressure on the supercell is in the order of 0.07 eV; similarly, if we assume the shear modulus has a value of 100 GPa, then the contribution of shear stress to the strain energy has a value on the order of 0.01 eV. These values are much smaller than the variations in core energies obtained in the calculations, which further verifies that the variance in core energies is not induced by the stress fluctuations in the supercells.

## a. Initial Configuration

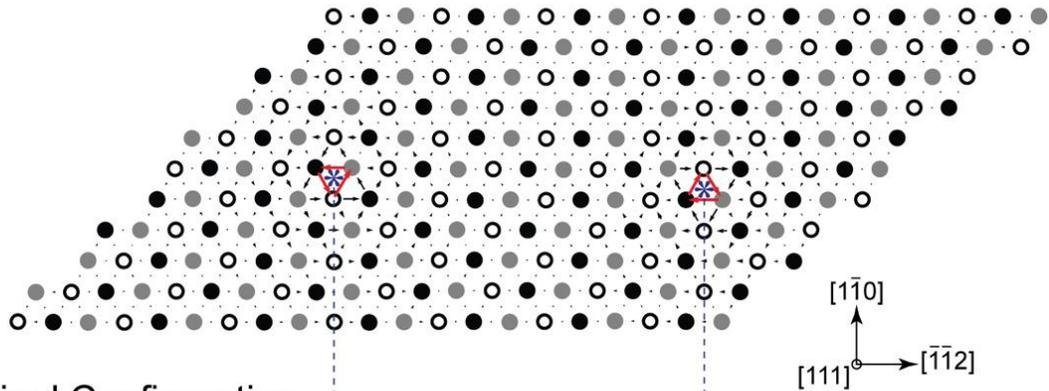

## b. Final Configuration

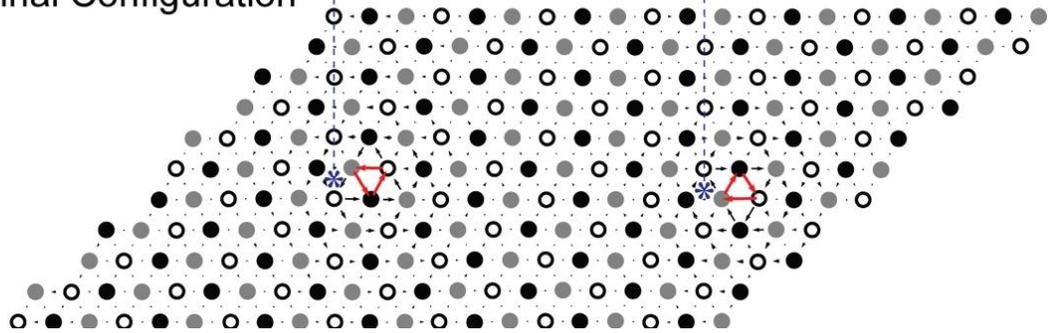

**Supplementary Figure 3 | Initial and final configurations of the dislocation dipole for the difference in Peierls valley calculations. a,** Initial configuration. **b,** Final configuration. Both configurations are plotted using differential displacement maps, and the compact dislocation core position is represented by the red arrows. For each relaxed sample selected as the initial configuration, as shown in **a**, we chose the sample with a nearest dislocation dipole on the same {110} plane and in the [$\bar{1}\bar{1}2$] direction as the final configuration.

**Supplementary Note 3: Calculation of diffuse antiphase boundary energy and its variation.**

In the dislocation dipole supercells, there is a cut plane between two dislocation cores, which will induce extra energy when SRO is present in the sample. The energy associated with the cut plane between dislocation cores varies depending on the level of chemical ordering and can be quantified through the diffuse antiphase boundary (DAPB) energy. In the following, we calculate the DAPB energy and its variance in our system for the *s1-s3* states in two different ways and based on the calculated DAPB energy, we quantify the contribution of the cut plane to the energy of the dislocation dipole supercell ($E_{DAPB}$).

Supplementary Figure 4 illustrates the first approach to calculating the averaged DAPB energies. Supplementary Figure 4a shows the supercell without dislocation dipole, in which six layers of atoms in the supercell are shifted by one Burgers vector in the [111] direction, as shown in Supplementary Figures 4b-c; accordingly, two DAPBs indicated by the blue dashed lines are created. The location of the DAPBs is then shifted in the [1$\bar{1}$0] direction, leading to 11 different configurations for each SRO state. All the configurations with DAPBs are then relaxed in the same way as described in the Method section and the average DAPB energies are calculated. The results are listed in Supplementary Table 3. For the state *s1* close to the random solid solution sample, $\gamma_{DAPB}$ is around 3 $mJ/m^2$, *i.e.*, essentially zero within the precision of the statistical sampling. With increasing SRO, the diffuse antiphase boundary energies increase; specifically, $\gamma_{DAPB}$ increases to 29 $mJ/m^2$ in *s2* and 59 $mJ/m^2$ in *s3*. With these averaged values, we can multiply $\gamma_{DAPB}$ by the area of the cut plan to obtain estimates of $E_{DAPB}$. These results are presented below and compared with values derived from a second approach that also provides insight into the variances of $E_{DAPB}$ with local atomic environment.

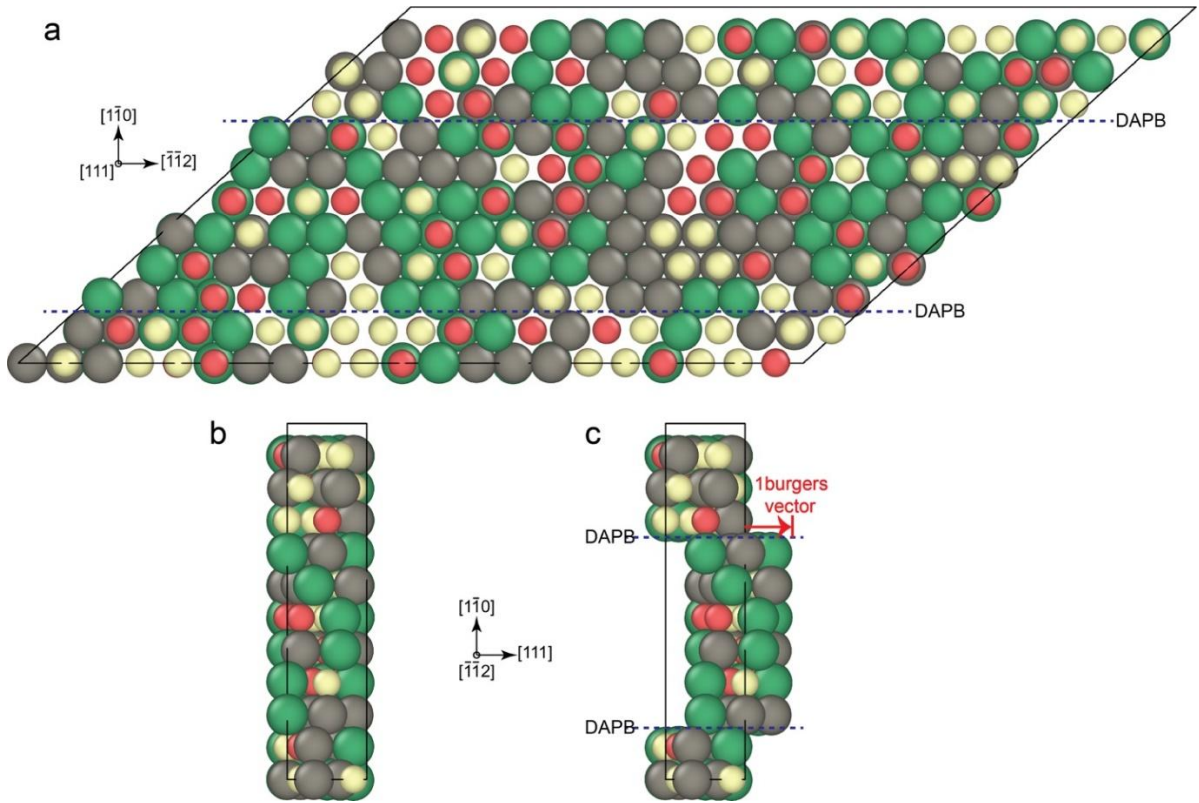

**Supplementary Figure 4 | Calculation of Diffuse Antiphase Boundary (DAPB) Energy. a,** Initial configuration without the dislocation dipole, the blue dashed lines represent two diffuse antiphase boundaries. **b,** Side view of the original supercell. **c,** Side view of the supercell with DAPBs. Six layers of atoms were shifted along [111] direction by 1 Burgers vector to create two DAPBs.

**Supplementary Table 3 | Diffuse Antiphase Boundary Energy.** Averaged DAPB energy for different SRO states.

|     | DAPB energy $\gamma_{DAPB}$ ($\frac{mJ}{m^2}$) | Standard error $\sigma_{\gamma_{DAPB}}/\sqrt{\# \ of \ samples}$ ($\frac{mJ}{m^2}$) |
| --- | --- | --- |
| *s1* | 3 | 2.2 |
| *s2* | 29 | 2.8 |
| *s3* | 59 | 2.7 |

To better mimic the effect of DAPB energy contribution, especially its variance in the dislocation dipole model, in the second approach we select a row of atoms as shown by the cyan dashed box in Supplementary Figure 5, which is the same length compared with the cut plane between two cores of a dislocation dipole. We shifted these atoms in [111] direction by one Burgers vector to mimic the effect of the cut plane. The system is then relaxed and the system energy $E_{DAPB\_segment}$ contains approximately double amount of the energy from the DAPB compare with the cut plane induced DAPB energy in a dislocation dipole model. Thus, the contribution of DAPB energy from the cut-plane in the dislocation dipole model is written as: $E_{DAPB} = \frac{E_{DAPB\_segment} - E_0}{2}$, where $E_0$ is the equilibrium energy of the cell without dislocation dipole and any DAPB and $E_{DAPB\_segment}$ is the equilibrium energy of the cell as shown in Supplementary Figure 5.

To quantify the variance of the $E_{DAPB}$ in a similar condition as the dislocation dipole energy shown in Figure 3, the cyan box illustrated in Supplementary Figure 5 is translated over all the positions in the simulation cell to create 231 different configurations for each SRO state. The average values of $E_{DAPB}$, the standard deviation $\sigma_{E_{DAPB}}$ and $\gamma_{DAPB}$ for three different SRO states are then calculated. The results are presented in Supplementary Table 4. The $\gamma_{DAPB}$ from this approach is very close to the value that we obtain from the first approach, given in Supplementary Table 3 from a much larger DAPB. If we take these values of $\gamma_{DAPB}$ and multiply by the area of the cut plane, we obtain estimates for $E_{DAPB}$ of 0.015 eV for *s1*, 0.284 eV for *s2* and 0.593 eV for *s3*.

Considering the excess supercell dipole energy shown in Figure 3, which is composed of core energies, elastic energy and the DAPB energy from the cut-plane ($E = 2E^{core} + E_{elastic} + E_{DAPB}$), we can decouple and calculate the contribution of variance due to the two dislocation cores and the diffuse antiphase boundary by $\text{Var}[2E^{core}]=\text{Var}[E]-\text{Var}[E_{DAPB}]$. The results are shown in Supplementary Table 4. The normalized core energy and its variance in this HEA system are calculated in the following section.

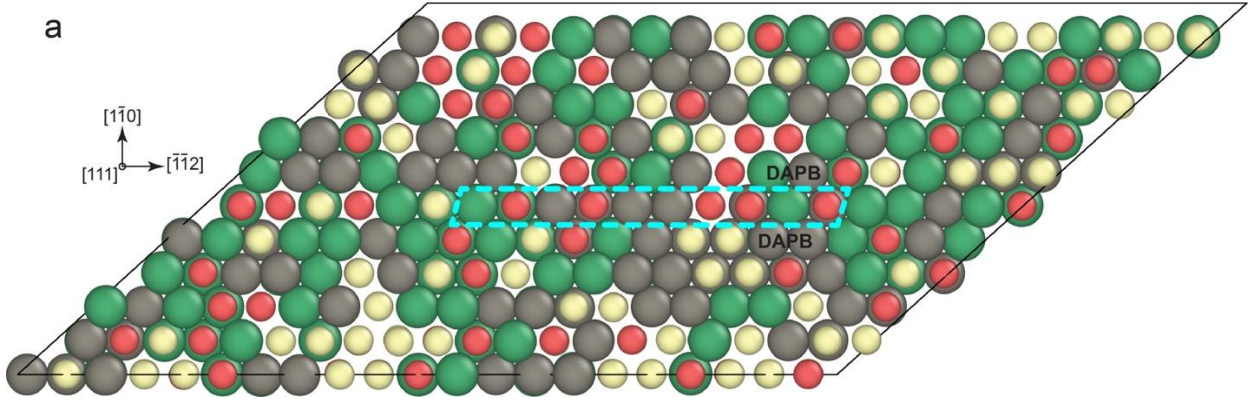

**Supplementary Figure 5 | Calculation of Diffuse Antiphase Boundary (DAPB) Energy and its variance. a,** The atoms in the cyan dashed box were shifted along [111] direction by one Burgers vector, which will lead the creation of two DAPB interfaces.

**Supplementary Table 4 | Diffuse Antiphase Boundary Energy.** Averaged DAPB energy for different SRO states.

|  | Average of $E_{DAPB}$ (eV) | Standard deviation of $\sigma_{E_{DAPB}}$ (eV) | DAPB energy $\gamma_{DAPB}$ ($\frac{mJ}{m^2}$) | Standard deviation of $\sigma_{2E^{core}}$ (eV) |
|---|---|---|---|---|
| s1 | 0.015 | 0.148 | 2 | 0.70 |
| s2 | 0.284 | 0.166 | 30 | 0.32 |
| s3 | 0.593 | 0.258 | 62 | 0.27 |

**Supplementary Note 4: Normalized core energy and its variance in HEAs.**

Considering the excess supercell dipole energy shown in Figure 3, which is composed of core energies, elastic energy and the DAPB energy from the cut-plane, we have the supercell excess energy $E$ in the form:

$$E = 2E_{core} + E_{elastic} + E_{DAPB} = 4b\hat{E}_{core} + E_{elastic} + E_{DAPB},$$

where $\hat{E}_{core}$ is the normalized core energy (units: meV/A).

Accordingly, the averaged value of $\hat{E}_{core}$ equals:

$$\hat{E}_{core} = (E - E_{elastic} - E_{DAPB})/4b.$$

The variance of $\hat{E}_{core}$ can also be calculated based on the variance of supercell excess energy $E$ and the variance of DAPB energy $E_{DAPB}$ (neglecting the elastic contribution).

Assuming that the dislocation dipole energy follows a normal distribution as shown in Figure 8a, we consider two dislocations in one dipole, termed *c1* and *c2*, with their core energies written as $E_{c1}$ and $E_{c2}$, which follow the same normal distribution, and we have $E_{c1} + E_{c2} = 2b\hat{E}_{core} + 2b\hat{E}_{core} = 2E_{core}$.

When considering the variance of the variables, we can write as follows:

$$Var[E] = Var[2E_{core}] + Var[E_{DAPB}] = Var[E_{c1}] + Var[E_{c2}] + Var[E_{DAPB}] = 2Var[2b\hat{E}_{core}] + Var[E_{DAPB}].$$

Thus,

$$Var[\hat{E}_{core}] = \frac{Var[E] - Var[E_{DAPB}]}{2 * 4b^2},$$

and:

$$\sigma_{\hat{E}_{core}} = \frac{1}{2b}\sqrt{\frac{Var[E] - Var[E_{DAPB}]}{2}}.$$

Based on the values of the supercell excess energy, the DAPB energy and elastic energy, discussed in the previous sections, we can obtain the normalized core energy and its variance; detailed values are provided in Supplementary Table 5. These data are also plotted in Figure 4.

**Supplementary Table 5 |** Normalized core energy and its standard deviation in the MoNbTaW system, compared with the core energy in pure *bcc* transition metals.

|  | Mo | Nb | Ta | W | Fe | HEA-*s1* | HEA-*s2* | HEA-*s3* |
|---|---|---|---|---|---|---|---|---|
| $\hat{E}_{core}$ (meV/A) | 416 | 171 | 153 | 501 | 206 | 492 | 504 | 503 |
| $\sigma_{\hat{E}_{core}}$ | - | - | - | - | - | 89 | 40 | 34 |

## Supplementary Note 5: Correlation and covariance between energy of neighboring dipoles.

Although the screw dislocation cores are high localized and compact, the energy of two neighboring dislocation dipoles, *i.e.*, two dipoles shown in Supplementary Figure 3, are not totally independent. Based on the relaxed energies of dislocation dipoles sampled over all the positions within the supercell for the three SRO states, the correlation coefficients between the energy of two neighboring dislocation dipole positions are calculated to be 0.85, 0.80, and 0.82 for *s1-s3* states, respectively. However, this correlation decays rapidly as the distance between dipole increases, as shown in Supplementary Figure 6.

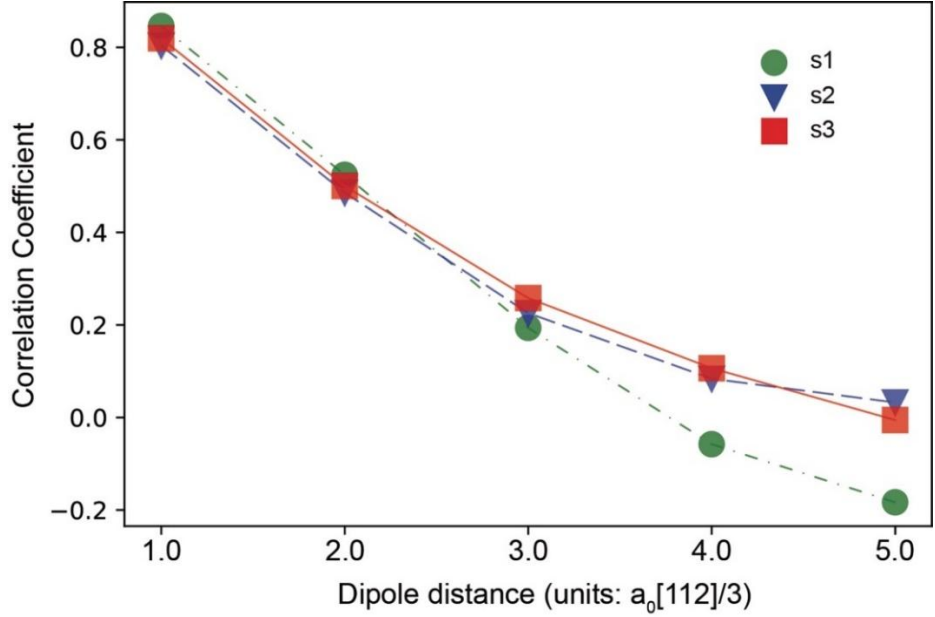

**Supplementary Figure 6 | Correlations coefficient of dipole energies as function of distance.**

Thus, when considering the distribution of Peierls valley energy differences in the main text, the covariance needs to be included in analyses of the distribution. Based on our calculations, the covariances of the two neighboring dipole energies are 0.44, 0.10, and 0.11 for *s1-s3* states, respectively. The predicted standard deviation ($\sqrt{2}\sqrt{\sigma^2 - \sigma_{cov}}$) for the Peierls valley energy difference equals 0.40, 0.24, and 0.23 eV, in excellent agreement with those obtained directly from the DFT data for *s1* to *s3,* as shown in Figure 6.

**Supplementary Table 6 | Correlation and covariance between energy of neighboring dipoles.**

|  | **Correlation Coefficient** | **Covariance $\sigma_{cov}$ ($eV^2$)** | $\sqrt{2}\sqrt{\sigma^2 - \sigma_{cov}}$ ($eV$) |
|---|---|---|---|
| *s1* | 0.85 | 0.44 | 0.40 |
| *s2* | 0.80 | 0.10 | 0.24 |
| *s3* | 0.82 | 0.11 | 0.23 |

**Supplementary Note 6: Analysis of Type-2 barrier for a single screw dislocation**

Assuming that the dislocation dipole energy follows a normal distribution: $Normal(\mu, \sigma^2)$, as shown in Figure 7a, we consider two dislocations in one dipole, termed *c1* and *c2*, with their core energies written as $E_{c1}$ and $E_{c2}$:

$$E_{c1} + E_{c2} \sim Normal(\mu, \sigma^2) \ .$$

Since two dislocation cores *c1* and *c2* are far away from each other, we can assume the core energies for these two dislocations are independent and their individual values also follow the same normal distribution. Under these assumptions, we can state that:

$$E_{c1} \sim Normal(\mu/2, \sigma^2/2), \ E_{c2} \sim Normal(\mu/2, \sigma^2/2).$$

Thus, a single dislocation follows the normal distribution: $Normal(\mu/2, \sigma^2/2)$.

Now if we consider two neighboring dislocations *w1* and *w2*, their core energies can be written as $E_{w1}$ and $E_{w2}$:

$$E_{w1} \sim Normal(\mu/2, \sigma^2/2), \ E_{w2} \sim Normal(\mu/2, \sigma^2/2) \ ,$$

and

$$E_{w1} - E_{w2} \sim Normal(0, \sigma^2 - 2\sigma_{w1w2}) \ .$$

where $\sigma_{w1w2}$ is the covariance of $E_{w1}$ and $E_{w2}$.

The criterion for a type-2 barrier is written as: $E_{w1} - E_{w2} > e_{critical}$; we can then write the probability of a Type-2 barrier as:

$$P_{type2} = P(E_{w1} - E_{w2} > e_{critical}) = 1 - P(E_{w1} - E_{w2} \leq e_{critical}) = 1 - \Phi(\frac{e_{critical}}{\sqrt{\sigma^2 - 2\sigma_{w1w2}}}),$$

where $\Phi$ is the standard Normal cumulative distribution function.